\documentstyle[12pt,citesort]{article}

\begin{document}

\title{{\bf Charge Fluctuations in Finite Coulomb Systems}}

\author{{\bf B.Jancovici}$^1$} 

\maketitle

\begin{abstract}
When described in a grand canonical ensemble, a finite Coulomb system
exhibits charge fluctuations. These fluctuations are studied in the case
of a classical (i.e. non-quantum) system with no macroscopic average 
charge. Assuming the validity of 
macroscopic electrostatics gives, on a three-dimensional finite
large conductor of volume $V$, a mean square charge $<Q^2>$ which goes
as $V^{1/3}$. More generally, in a short-circuited capacitor of
capacitance $C$, made of two conductors, the mean square charge on one
conductor is $<Q^2>=TC$, where $T$ is the temperature and $C$ the
capacitance of the capacitor. The case of only one conductor in a
grand canonical ensemble is obtained by removing the other conductor to
infinity. The general formula is checked in the weak-coupling
(Debye-H\"uckel) limit for a spherical capacitor. For two-dimensional
Coulomb systems (with logarithmic interactions), there are exactly
solvable models which reveal that, in some cases, macroscopic
electrostatics is not applicable even for large conductors. This is when
the charge fluctuations involve only a small number of particles. The
mean square charge on one two-dimensional system alone, in the grand
canonical ensemble, is expected to be, at most, one squared elementary 
charge. 
\end{abstract}

\medskip

\noindent {\bf KEY WORDS:} Finite Coulomb systems; charge fluctuations.

\medskip

\noindent LPT Orsay 01-112

\vfill

\noindent $^1$Laboratoire de Physique Th\'eorique, B\^atiment 210, 
Universit\'e de Paris-Sud, 91405 Orsay, France (Unit\'e Mixte de 
Recherche no.8627-CNRS); e-mail: Bernard.Jancovici@th.u-psud.fr

\newpage

\renewcommand{\theequation}{1.\arabic{equation}}
\setcounter{equation}{0}

\section{\bf INTRODUCTION}

This paper is dedicated to Michael Fisher on the occasion of his 70th
birthday. Part of it deals with the Debye-H\"uckel theory of Coulomb
systems, to which Michael brought elaborate refinements, in particular
insisting on the importance of taking hard cores into account. I 
apologize for using here only the simple point-particle version. 

I consider the classical (i.e. non-quantum) equilibrium statistical
mechanics of Coulomb systems: systems of charged particles interacting
through the Coulomb law (plus perhaps short-ranged forces), such as
plasmas or electrolytes. Such a system, when described by a grand
canonical ensemble, is expected to exhibit charge fluctuations. The aim
of the present paper is to study these fluctuations, for a finite but
macroscopic system. 

The grand canonical ensemble is often introduced by considering a
system in contact with a surrounding infinite 
system (the reservoir), with possible exchanges of energy and particles
between the system and the reservoir. In this approach, it is assumed 
that the energy of interaction between the system and the reservoir is
negligible. This is indeed the case for a finite but macroscopic system,
when the interparticle forces are short-ranged (then, the system energy
goes as its volume, while the interaction energy goes only as its
surface area). However, this approach has to be modified when there
are long-ranged forces such as Coulomb ones. Then, for the interaction 
energy between the system and the reservoir be disregarded, is is
necessary to assume that the reservoir is infinitely far away from
the system under consideration. This is how the grand canonical ensemble
will be defined in the following.        
 
Studying the charge fluctuations in a given large subvolume $\Lambda$ of 
an infinite Coulomb system (i.e. assuming that the reservoir is in
contact with the subvolume) is a different problem, which has already
been studied and solved\cite{Martin,Levesque}. It was found that the 
mean square charge $\langle Q^2\rangle$ in $\Lambda$ behaves as its 
surface area $S$ (not its volume $V$). In the presently studied geometry 
(infinitely remote reservoir), it will be argued that 
$\langle Q^2\rangle$  is even weaker, 
behaving as $V^{1/3}$, for a 3-dimensional system. This behavior had
been conjectured by Lieb and Lebowitz\cite{Lieb}, but they could not
prove it rigorously, i.e. by a purely microscopic argument. Here, on 
the contrary, our starting point will be the validity of macroscopic 
electrostatics of conductors, assumed without proof.

In general, in the grand canonical ensemble, the state of a system, 
made of $s$ species of particles, depends on $s$ chemical
potentials. However, in the case of a Coulomb system, 
in the thermodynamic limit, the state depends only on $s-1$
chemical potentials and is the same as in a grand canonical ensemble
restricted to neutral configurations\cite{Lieb}. In the case of a
macroscopic but finite Coulomb system, there may be a non-zero average 
charge $\langle Q\rangle$ and one more chemical potential is needed for
controlling it. Here, we only consider the simple case in which there is
only one reservoir, which is an infinite Coulomb  system of the same
nature as the system under consideration, and $\langle Q\rangle$ is
expected to vanish if macroscopic electrostatics holds.

We shall first assume that our system and this reservoir are at some
distance of each other. Since they can freely exchange particles, the 
system and the reservoir can be considered as two conductors forming a 
short-circuited capacitor. In Section 2, a general simple
expression for the mean square charge of the system in that
configuration will be derived; the case of the system in the
grand canonical ensemble will be obtained by removing the reservoir to
infinity. In Section 3, these general results will be checked by a
microscopic calculation in the weak-coupling (Debye-H\"uckel) limit. 

Coulomb systems can be mimicked in a two-dimensional world,
in which the Coulomb potential $1/r$ must be replaced by the
two-dimensional solution of the Poisson equation $-\ln r$. Working in
two dimensions has the advantage that exactly solvable models for the 
statistical mechanics of Coulomb systems are available. However, in two 
dimensions, some specific subtle points arise and deserve a separate
discussion. In Section 4, two-dimensional models will be considered, 
and cases when macroscopic electrostatics cannot be used will be 
exhibited. 

Rather than starting with a capacitor made of two conductors and
removing one of them to infinity, one might want to study directly the
case of one finite system in the grand canonical ensemble. How to
formulate the Debye-H\"uckel theory in a finite system is discussed
in Appendix B.

\renewcommand{\theequation}{2.\arabic{equation}}
\setcounter{equation}{0} 
  
\section{\bf CHARGE FLUCTUATIONS IN A CAPACITOR}

\hspace*{\parindent} Let us consider a capacitor, made of two
macroscopic conductors $A$ and $B$ separated by vacuum, with $B$
surrounding $A$. Let $C$ be the capacitance. Let $Q$ be the
charge on conductor $A$, $-Q$ the charge on 
conductor $B$. When the capacitor is short-circuited, the average 
charge $\langle\pm Q\rangle $ on each conductor vanishes. 
If the conductors are describable by classical (non-quantum) 
statistical mechanics, we claim that the mean square charge on each 
conductor is 
\begin{equation} \label{2.1}
\langle Q^2\rangle =T C 
\end{equation}
where $T$ is the temperature (in units of energy). 
This relation is a special case of Nyquist's formula\cite{Landau} which
gives the electrical current fluctuations in a linear electric circuit. 
The derivation of Nyquist's formula relies on the
fluctuation-dissipation theorem in the theory of linear response. In the
present case, a more direct derivation of eq.(\ref{2.1}) can be made
by using the simpler classical static linear response theory, as follows.

Let us introduce some given external charges which create an
infinitesimal electric potential difference between the two conductors:
let the potential on conductor $A$ minus the potential on conductor 
$B$ be $\delta\phi$ (for instance, in the case of a spherical capacitor,
we introduce between the conductors two spherical concentric layers 
carrying opposite uniformly distributed charges; then, these layers 
create a potential which is zero on the outer conductor $B$ and has 
some constant value $\delta\phi$ on the inner conductor $A$). The 
corresponding change in the Hamiltonian is $H'=\delta\phi\,Q$, where $Q$ 
is the charge on conductor $A$. The short-circuited capacitor will
respond by tranfering a charge $\delta Q =-C\,\delta\phi$ from conductor
$B$ to conductor $A$, in such a way that the total potential difference 
(external plus induced) between the conductors vanishes. Now, linear 
response theory says
\begin{equation} \label{2.2}
-C\,\delta\phi = \delta Q = -\beta (\langle H'Q\rangle -\langle H'
\rangle\langle Q\rangle) = -\beta\,\delta\phi\langle Q^2\rangle
\end{equation}
where $\beta$ is the inverse temperature $1/T$ and the averages 
$\langle\ldots\rangle $ are taken in the absence of the perturbation
$H'$ (we have used $\langle Q\rangle =0$). Eq.(\ref{2.2}) proves 
eq.(\ref{2.1}).It should be noted that $C$ will be the usual capacitance
for a given geometry only provided that macroscopic electrostatics is 
applicable. A necessary condition is that the sizes of the conductors 
and the separation between them be large compared to the microscopic
scale.

The mean square charge of a finite macroscopic Coulomb system in the
grand canonical ensemble is obtained by sending conductor $B$ to
infinity. Then the mean square charge on the macroscopic body $A$ 
is given by (\ref{2.1}), where now $C$ is the capacitance of the 
macroscopic body $A$ alone. For a 3-dimensional system, this capacitance
goes as $V^{1/3}$ where $V$ is the volume, as stated in the
Introduction.

In the special simple case of a spherical capacitor, made of an inner 
conductor of radius $R_1$ and an outer conductor of radius $R_2$, the 
capacitance is
\begin{equation} \label{2.3}
C = \frac{1}{(1/R_1)-(1/R_2)}
\end{equation}
It becomes $R_1$ if $R_2\rightarrow\infty$, giving the mean square
charge in a macroscopic but finite spherical Coulomb system of radius
$R$, in the grand-canonical ensemble,
\begin{equation} \label{2.4}
\langle Q^2\rangle =TR
\end{equation}
in agreement with previous findings about the charge correlations.
\cite{BJ}

Another limit of interest is when $R_1$ becomes large
compared to $W=R_2-R_1$ which keeps a fixed (macroscopic) value. One
then obtains a plane capacitor with plate areas $S=4\pi R_1^2$ and plate
separation $W$, and indeed (\ref{2.3}) becomes $C = S/(4\pi W)$,
giving for the mean square charge per unit area on one plate 
\begin{equation} \label{2.5}
\lim_{S\rightarrow\infty}\frac{\langle Q^2\rangle }{S} =\frac{T}{4\pi W}
\end{equation}

\renewcommand{\theequation}{3.\arabic{equation}}
\setcounter{equation}{0}

\section{\bf WEAK-COUPLING LIMIT}

\hspace*{\parindent} We consider the simple geometry of a spherical
capacitor made 
with a classical Coulomb fluid: There are two concentric spheres,
centered at the origin, of radii $R_1$ and $R_2 \; (R_2>R_1)$. The 
shell between the spheres is empty. The  ball of radius $R_1$ is
filled with the fluid, as well as the whole space outside the sphere of
radius $R_2$. The short-circuiting of the capacitor is described by 
assuming that the two filled regions are allowed to freely exchange 
particles.

The weak-coupling limit is a high-temperature one which is expected to 
be correctly described by the Debye-H\"uckel theory. The Coulomb fluid
is made of several species of particles of number densities $n_a$ and 
charges $q_a$. For the system to be stable, in addition to the Coulomb 
forces, there should be some short-ranged repulsive forces, but the 
weak-coupling limit can also be viewed as a low-density one in which  
these short-ranged forces can be neglected. Strictly speaking, the
number densities are position-dependent near the fluid boundaries. 
However, taking this into account 
would only give corrections of higher order and therefore we consider 
the densities as constants. The Debye 
wave number is defined as $\kappa =(4\pi \beta \sum_a n_aq_a^2)^{1/2}$. 
Let $\rho({\mathbf r})$ be the microscopic charge density at
${\mathbf r}$. We shall need the charge correlation function  
\begin{equation} \label{3.1}
\langle\rho ({\mathbf r})\rho ({\mathbf r'})\rangle =\sum_{ab}
n_aq_a^2n_bq_b^2K({\mathbf r},{\mathbf r'})+\sum_an_aq_a^2
\delta({\mathbf r}-{\mathbf r'})
\end{equation}
where both ${\mathbf r}$ and ${\mathbf r'}$ are in a filled region.   
$K({\mathbf r},{\mathbf r'})$ is the solution of the partial 
differential equation
\begin{equation} \label{3.2}
[\Delta -\kappa^2({\mathbf r})]K({\mathbf r},{\mathbf r'})=
4\pi\beta\delta({\mathbf r}-{\mathbf r'})
\end{equation}
where the source point ${\mathbf r'}$ is assumed to be in a filled 
region, while ${\mathbf r}$ can be anywhere: 
$\kappa^2({\mathbf r})=\kappa^2$ if ${\mathbf r}$ is in a
filled region and $\kappa^2({\mathbf r})=0$ if ${\mathbf r}$ is in the
empty region. $K$ and its normal derivative must be continuous at the
boundaries $r=R_1$ and $r=R_2$, and $K\rightarrow 0$ as
$r\rightarrow\infty$.

In the present Debye-H\"uckel scheme, the average charge on the sphere 
of radius $R_1$ vanishes, while the mean square charge on that sphere is
\begin{equation} \label{3.3} 
\langle Q^2\rangle =\int_{r<R_1}{\mathrm d}{\mathbf r}\int_{r'<R_1}
{\mathrm d}{\mathbf r'}\langle\rho ({\mathbf r})\rho 
({\mathbf r'})\rangle
\end{equation}
The solution $K$ of (\ref{3.2}) is studied in Appendix A. When used in 
(\ref{3.1}) and (\ref{3.3}) it gives
\begin{equation} \label{3.4}
\beta\langle Q^2\rangle =\frac{[1+\kappa R_2][\kappa R_1
\cosh (\kappa R_1)-\sinh (\kappa R_1)]}{\kappa [(1+\kappa R_2-\kappa
R_1)\cosh (\kappa R_1)+\sinh (\kappa R_1)]}
\end{equation}
If both $R_1$ and $R_2$ are macroscopic, i.e. if $\kappa R_1,\kappa R_2
\gg 1$, (\ref{3.4}) becomes
\begin{equation} \label{3.5}
\beta\langle Q^2\rangle =\frac{\kappa R_1R_2}{2+\kappa R_2-\kappa R_1}
\end{equation}
 
The mean square charge in a large spherical subdomain of an
infinite Coulomb system is retrieved by taking $R_1=R_2=R$ in
(\ref{3.5}) which becomes
\begin{equation} \label{3.6}
\beta\langle Q^2\rangle =\frac{1}{8\pi }\kappa S
\end{equation}
where here $S=4\pi R^2$ is the sphere area, in agreement with the
general formula\cite{Martin}
\begin{equation} \label{3.7}
\frac{\langle Q^2\rangle }{S}=-\frac{1}{4}\int {\mathrm d}
{\mathbf r}\,r\langle\rho (0)\rho ({\mathbf r})\rangle
\end{equation}
where $\langle\rho (0)\rho ({\mathbf r})\rangle $ is the infinite-system 
charge correlation function, obtained by using in (\ref{3.2}) the
infinite-system function $K(0,{\mathbf r})=-\beta\exp (-\kappa r)/r$.

If, on the contrary, $R_2-R_1$ is also macrosopic, i.e if we now assume
$\kappa (R_2-R_1)\gg 1$ as well as $\kappa R_1\gg 1$ in (\ref{3.5}), 
we do check the expected spherical capacitor charge fluctuation 
(\ref{2.1}) with capacitance (2.3), as well as the grand-canonical
fluctuation (\ref{2.4}). The limit (\ref{2.5}) of a plane capacitor can 
also be taken.

The plane capacitor can also be studied directly. There are two parallel
planes $x=0$ and $x=W$, thus  separated by a distance $W$. The slab 
between the plates $0<x<W$ is empty, while the Coulomb fluid fills the 
two semiinfinite regions $x<0$ and $x>W$ outside the slab. In this
geometry, it is possible to solve (\ref{3.2}) for a function $K$ which
now is a function of $x$, $x'$, and the component ${\mathbf y}$ of 
${\mathbf r}-{\mathbf r'}$ along the plates (actually, one rather
computes the Fourier transform of $K$ with respect to ${\mathbf y}$).
Then, one obtains the mean square charge per unit area on one plate of
(infinite) area $S$ as
\begin{equation} \label{3.8}
\lim_{S\rightarrow\infty}\frac{\langle Q^2\rangle }{S}=
\int {\mathrm d}{\mathbf y}\int_{x<0}{\mathrm d}x\int_{x'<0}
{\mathrm d}x'\langle\rho (x,{\mathbf y})\rho (x',0)\rangle
\end{equation}
The result is
\begin{equation} \label{3.9}
\lim_{S\rightarrow\infty}\frac{\beta\langle Q^2\rangle }{S}=
\frac{\kappa}{4\pi (2+\kappa W)}
\end{equation}
If $W=0$, one retrieves (\ref{3.6}). If $W$ is macroscopic, i.e.
$\kappa W\gg 1$, one retrieves (\ref{2.5}).
       
The above considerations also apply to the case of a one-component
plasma (also called jellium), i.e. a system made of mobile point 
charges of one species, with number density $n$ and charge $q$, in a 
uniform neutralizing background. The Debye wave number reduces to 
$\kappa=(4\pi\beta nq^2)^{1/2}$. However, for classical jellium, the
grand canonical partition function is a convergent series only if the
background is kept fixed while one sums over the particle number
\cite{LN}. The same kind of prescription should be used here: regions
$A$ and $B$ are assumed allowed to exchange particles, but the
backgrounds keep a fixed charge density. 

\renewcommand{\theequation}{4.\arabic{equation}}
\setcounter{equation}{0}

\section{\bf TWO-DIMENSIONAL COULOMB SYSTEMS}

\subsection{\bf General Properties}

\hspace*{\parindent} In the two-dimensional systems discussed in this 
Section, the electric potential created at ${\mathbf r}$ by a unit 
charge at the 
origin is $-\ln (r/L)$ where $L$ is some fixed length. This choice of a
two-dimensional solution of the Poisson equation often makes these
systems good toy models for mimicking three-dimensional systems with 
the usual $1/r$ potential. One of the advantages of working in two
dimensions is the existence of exactly solvable models. For avoiding any
confusion, it should be stressed that these toy models do \emph{not}
describe ``real'' charged particles such as electrons, which, even when 
confined in a surface, still interact by the $1/r$ law.

In the simple case of a circular capacitor, made of an inner circular
conductor of radius $R_1$ and an outer circular conductor of radius
$R_2$, macroscopic two-dimensional electrostatics says that the 
capacitance is 
\begin{equation} \label{4.1}
C=\frac{1}{\ln\frac{R_2}{R_1}}
\end{equation}
$C$ goes to 0 as $R_2\rightarrow\infty$.\footnote[2]{In a previous 
paper\cite{BJ}, another definition of the capacitance of a disk was
used. The present one (the limit of the capacitance of a circular
capacitor when the outer conductor recedes to infinity) is more
appropriate here.} More generally, the macroscopic capacitance of one 
finite conductor of any shape vanishes. This is because bringing from 
infinity an additional charge $q$ onto a conductor of characteristic 
size $R$, already carrying a charge $Q$, would cost an energy of order 
$qQ\int_R^{\infty}{\mathrm d}r/r$ which is infinite.  

It will however be shown below that, for the present problem of charge
fluctuations, the mimicking of three-dimensional systems which might be
expected at first glance does \emph{not} always occur, because, in some 
cases, although all relevant lengths are macroscopic, nevertheless
macroscopic electrostatics cannot be used. In the case of a circular
capacitor, $\langle Q^2\rangle $ may not be given by using (\ref{4.1})
in (\ref{2.1}). 

\subsection{\bf Weak-Coupling Limit}

\hspace*{\parindent} The two-dimensional case is very similar to the 
three-dimensional one. 
We now consider the simple geometry of a circular capacitor replacing
the concentric spheres by concentric circles: The regions $r<R_1$ and
$r>R_2$ are occupied by the Coulomb fluid, while the region $R_1<r<R_2$
is empty.  The Debye-H\"uckel theory 
is again used, with $4\pi$ replaced by $2\pi$ in the definition of the
Debye wave number $\kappa$ and in the r.h.s. of (\ref{3.2}). The detail
of the calculation is given in Appendix A. The result is
\begin{equation} \label{4.2}
\beta\langle Q^2\rangle =\frac{\kappa R_1}{\frac{I_0(\kappa R_1)}
{I_1(\kappa R_1)}+\frac{R_1}{R_2}\frac{K_0(\kappa R_2)}{K_1(\kappa R_2)}
+\kappa R_1\ln\frac{R_2}{R_1}}
\end{equation}
where $I_l$ and $K_l$ are modified Bessel functions, while $\langle Q
\rangle =0$. If both $R_1$ and $R_2$ are macroscopic, i.e. if 
$\kappa R_1,\kappa R_2 \gg 1$, 
(\ref{4.2}) becomes
\begin{equation} \label{4.3}
\beta\langle Q^2\rangle =\frac{\kappa R_1}{1+\frac{R_1}{R_2}+\kappa R_1
\ln\frac{R_2}{R_1}}
\end{equation}

The mean square charge in a large circular subdomain of an infinite
system is retrieved by taking $R_1=R_2=R$ in (\ref{4.3}) which
becomes 
\begin{equation} \label{4.4}
\beta\langle Q^2\rangle =\frac{\kappa}{4\pi}S
\end{equation}
where $S=2\pi R$ is the subdomain perimeter, in agreement with the
two-dimensional analog\cite{Martin} of (\ref{3.7}):
\begin{equation} \label{4.5} 
\frac{\langle Q^2\rangle }{S}=-\frac{1}{\pi}\int {\mathrm d}
{\mathbf r}\,r\langle\rho (0)\rho ({\mathbf r})\rangle
\end{equation}
where $\langle\rho (0)\rho ({\mathbf r})\rangle $ is the infinite-system
charge correlation function, obtained by using in (\ref{3.2}) the 
infinite-system function $K(0,{\mathbf r})=-\beta K_0(\kappa r)$.

If, on the contrary, $R_2-R_1$ is also macroscopic, i.e. if $\kappa 
(R_2-R_1) \gg 1$ as well as $\kappa R_1 \gg 1$, (\ref{4.3}) becomes
\begin{equation} \label{4.6}
\beta\langle Q^2\rangle =\frac{1}{\ln\frac{R_2}{R_1}}
\end{equation}
in agreement with (\ref{2.1}) and (\ref{4.1}). In the limit
$R_2\rightarrow\infty$, there is no charge fluctuation.

\subsection{\bf Two-Component Plasma at $\Gamma=2$}

\hspace*{\parindent} The two-dimensional two-component plasma is made of
two species  
of point particles of opposite charges $\pm q$, interacting through the
pair potential $\pm q^2\ln(r/L)$. The dimensionless coupling constant is
$\Gamma=\beta q^2$. The system is stable against collapse of
pairs of oppositely charged particles for $\Gamma <2$. Many exact
results are now available for that model in its whole stability
range\cite{ST,SJ1,S,SJ2,SJ3}. However, the correlation functions are
fully known only at $\Gamma=2$, for a variety of 
plane\cite{G,CJ1,CJ2,PJF,JS,GT} (or even curved) geometries (although, 
for a given fugacity, the density starts to diverge at $\Gamma=2$, the
truncated many-body distribution functions remain finite). 
 
By an extension of previous calculations, the correlation functions at
$\Gamma=2$, in the circular capacitor geometry described in Section 4.2 
are obtained in Appendix C. The fugacity $z$ (which is the same for
both species) appears through a parameter $m=2\pi Lz$ which has the
dimension of an inverse length and defines a microscopic scale (the
bulk correlation length is $(2m)^{-1}$). Using these correlation
functions in (\ref{3.3}) gives 
\begin{equation} \label{4.7}
\langle Q^2\rangle =q^2\sum_{l=-\infty}^{\infty}\frac{[(X^2+l^2)I_l^2(X)
-X^2I_l^{\prime 2}(X)][lK_l^2(\alpha X)-\alpha XK_l(\alpha X)K'_l(\alpha X)]}
{[lI_l(X)K_l(\alpha X)(\alpha^l-\alpha^{-l})-
XI_l(X)K'_l(\alpha X)\alpha^{l+1}+XI'_l(X)K_l(\alpha X)\alpha^{-l}]^2}
\end{equation}
where $X=mR_1$ and $\alpha=R_2/R_1$. By charge symmetry, 
$\langle Q\rangle =0$.

The mean square charge in a large circular subdomain of radius $R$ of an 
infinite system is retrieved by taking $X=mR$ and $\alpha=1$ in
(\ref{4.7}) and using for the modified Bessel functions $I_l$ and $K_l$
the leading terms of their uniform asymptotic expansions\cite{AS}, which
are appropriate in the present case of a large argument $X$ and an index
$l$ which may also be large. It is found that the sum on $l$ can be
replaced by an integral. The result is
\begin{equation} \label{4.8}  
\langle Q^2\rangle =q^2\frac{m}{8}S
\end{equation}
where $S=2\pi R$ is the subdomain perimeter, in agreement with the 
general formula (\ref{4.5}), where the infinite-system charge
correlation function\cite{CJ1,CJ2} 
$\langle\rho(0)\rho({\mathbf r})\rangle $ is (up to its here irrelevant
$\delta$ term) $-2q^2[m^2/(2\pi)]^2[K_0^2(mr)+K_1^2(mr)]$.

If, on the contrary, not only $R_1$ but also $R_2-R_1$ are macroscopic,
i.e. $X\gg 1$ with $\alpha >1$, because of the $\alpha^l,\alpha^{-l},
\alpha^{l+1}$ terms in the denominator of (\ref{4.7}) only small values
of $|l|$ contribute to the sum, and the modified Bessel functions can be
simply replaced by the leading terms of their ordinary asymptotic
expansions\cite{AS}, valid for a large argument $X$ or $\alpha X$ and a 
given index $l$. In the limit $X=mR_1\rightarrow\infty$ at fixed 
$\alpha =R_2/R_1$, with $\beta q^2=2$ being taken into account, 
(\ref{4.7}) becomes
\begin{equation} \label{4.9}
\beta\langle Q^2\rangle =2\sum_{l=-\infty}^{\infty}\frac{1}
{\left(\alpha^{l+\frac{1}{2}}+\alpha^{-(l+\frac{1}{2})}\right)^2}
\end{equation}   
where $\alpha =R_2/R_1$. It is clear that, when $\alpha$ has a finite
value, the sum in (\ref{4.9}) cannot be replaced by an integral [which
would reproduce (\ref{4.6})].

(\ref{4.9}) is, at first sight, a very surprising result. It does
\emph{not} reproduce the value (\ref{4.6}) expected on the basis of
macroscopic electrostatics. On second thought, one sees that the l.h.s. 
of (\ref{4.9}) can be rewritten as $2\langle Q^2\rangle /q^2$ since 
here $\Gamma=\beta q^2=2$, and therefore $\langle Q^2\rangle /q^2$ is 
of order unity, which means that the fluctuation involves only a small 
number of particles. Thus, in spite of the fact that the relevant 
lengths $R_1$ and $R_2-R_1$ are macroscopic, the number of involved 
particles is not, and (\ref{4.6}) based on macroscopic electrostatics 
should not be expected to hold at $\Gamma=2$, and more generally at any
finite temperature. However, in the weak-coupling
(i.e. high-temperature) limit considered in 
Section 4.2, $\beta q^2\rightarrow 0$, $\langle Q^2\rangle /q^2$ as
given by (\ref{4.6}) becomes large, and this result is consistent 
with macroscopic electrostatics, as it should.

It should be noted that, like (\ref{4.6}), (\ref{4.9}) indicates that 
there is no charge fluctuation in the limit $R_2\rightarrow\infty$,
i.e. for one macroscopic disk in the grand canonical ensemble. Only the
charge $Q=0$ contributes to the grand canonical distribution.
  
Another limit of interest is when $R_1\rightarrow\infty$ for a fixed
value of $W=R_2-R_1$. One then obtains a two-dimensional ``plane'' 
capacitor. When this limit is approached, $\ln\alpha\sim W/R_1$ is
small, and the sum in (\ref{4.9}) can be replaced by the integral
\begin{equation} \label{4.10}
\beta\langle Q^2\rangle =\frac{1}{2}\int_{-\infty}^{\infty}
\frac{{\mathrm d}l}{\cosh^2(lW/R_1)}=\frac{2\pi R_1}{2\pi W}
\end{equation}
This is the result expected on the basis of macroscopic electrostatics,
the two-dimensional analog of (\ref{2.5}) with now the plate area
replaced by a plate length $2\pi R_1$ and the capacitance the
two-dimensional one $C=2\pi R_1/(2\pi W)$. Since $\langle Q^2\rangle
/q^2$ now becomes large as $R_1\rightarrow\infty$, macroscopic 
electrostatics should indeed hold. A direct derivation for a plane 
capacitor is feasible.
    
\subsection{\bf One-Component Plasma at $\Gamma =2$}

\hspace*{\parindent} The two-dimensional one-component plasma is made of
one species 
of point particles of charge $q$, interacting through the pair potential
$-q^2\ln(r/L)$, in a neutralizing background of fixed charge density
$-qn$. Far from the boundaries, the particle number density is $n$. 
The dimensionless coupling constant again is $\Gamma =\beta q^2$. 
Up to now, the system is exactly solvable\cite{AJ} only at 
$\Gamma =2$, in which case the correlation functions are known for a 
large variety of plane\cite{BJ1,BJ2,CJB,BJ3} (or even curved) 
geometries. 

By an extention of previous calculations, the charge density and the 
correlation function at $\Gamma =2$, in the circular capacitor
geometry, are obtained in Appendix D. The background is fixed, while the
two regions can freely exchange particles. There is no charge symmetry
and the average charge $\langle Q\rangle $ on the inner disk does not 
automatically vanish.  Using in (\ref{3.3}) (modified for taking 
$\langle Q\rangle$ into account), the correlation function of Appendix D 
gives the charge fluctuation on the inner disk. We introduce the
notations $Y_1=\pi nR_1^2$ and $Y_2=\pi nR_2^2$ and the incomplete gamma
functions\cite{E}
\begin{equation} \label{4.11}
\gamma(l+1,Y_1)=\int_0^{Y_1}{\mathrm d}t\,{\mathrm e}^{-t}t^l
\end{equation}
and
\begin{equation} \label{4.12}
\Gamma(l+1+Y_2-Y_1,Y_2)=\int_{Y_2}^{\infty}{\mathrm d}t\,
{\mathrm e}^{-t}t^{l+Y_2-Y_1}
\end{equation}
The average charge on the inner disk is found to be
\begin{equation} \label{4.13}
\langle Q\rangle =q\sum_{l=0}^{\infty}\frac{\gamma (l+1,Y_1)}
{\gamma (l+1,Y_1)+D(l+1,Y_1,Y_2)}-qY_1
\end{equation}
where
\begin{equation} \label{4.14}
D(l+1,Y_1,Y_2)=\exp(Y_1\ln Y_1-Y_1-Y_2\ln Y_2+Y_2)\Gamma(l+1+Y_2
-Y_1,Y_2)
\end{equation}
The charge fluctuation on the inner disk is found to be
\begin{equation} \label{4.15}
\langle Q^2\rangle -\langle Q\rangle^2=q^2\sum_{l=0}^{\infty}\frac
{\gamma(l+1,Y_1)D(l+1,Y_1,Y_2)}{[\gamma(l+1,Y_1)+D(l+1,Y_1,Y_2)]^2}
\end{equation}
The integrands in (\ref{4.11}) and (\ref{4.12}) have a maximum (a sharp
one when $l$ is large) at $t=l$ and $t=l+Y_2-Y_1$, respectively. 
Therefore, $\gamma(l+1,Y_1)$ becomes small when $l>Y_1$, while 
$\Gamma(l+1+Y_2-Y_1,Y_2)$ and thus $D(l+1,Y_1,Y_2)$ become small 
when $l<Y_1$. The summand in (\ref{4.15}) has a maximum 
near $l=Y_1$.

The results for a large circular subdomain of radius $R$ of an
infinite system are retrieved by taking $R_1=R_2=R$, i.e 
$Y_1=Y_2=\pi nR^2$. Then (\ref{4.13}) gives $\langle Q\rangle =0$.
Using for the incomplete gamma functions
(\ref{4.11}) and (\ref{4.12}) the Tricomi asymptotic 
representation\cite{E} 
\begin{equation} \label{4.16}
\gamma(l+1,Y)\sim\Gamma (l+1)[\frac{1}{2}+\pi^{-1/2}{\mathrm Erf}
(\frac{Y-l}{(2Y)^{1/2}})]
\end{equation}
where $\Gamma$ is the complete gamma function and Erf is the error 
function (this representation is
appropriate when $Y$ is large and $l$ is close to $Y$), it is 
found that the sum on $l$ in (\ref{4.15}) can be replaced by an
integral. The result is
\begin{equation} \label{4.17}
\langle Q^2\rangle =q^2\frac{n^{1/2}}{2\pi }S
\end{equation}
where $S=2\pi R$ is the subdomain perimeter, again in agreement with the
general formula (\ref{4.5}), where the infinite-system charge 
correlation function\cite{BJ1} $\langle\rho(0)\rho({\mathbf r})\rangle $ 
now is (up to its here irrelevant $\delta$ term)  
$-q^2n^2\exp(-\pi nr^2)$.

We now turn to the case when not only $R_1$ but also $R_2-R_1$ are 
macroscopic, i.e. $Y_1\gg 1$ with $\alpha =R_2/R_1=(Y_2/Y_1)^{1/2}>1$.
The average charge (\ref{4.13}) does not seem to have a simple 
expression. For investigating the charge fluctuation (\ref{4.15}), it is
convenient to rewrite the summand in it as 
$[(D/\gamma)^{1/2}+(\gamma/D)^{1/2}]^{-2}$. As it will be seen
a posteriori, now the relevant values of $l-Y_1$ in the sum (\ref{4.15}) 
are only of the order of a few units (rather than of the order of
$Y_1^{1/2}$)  and therefore the Erf term in the Tricomi representation 
(\ref{4.16}) can be omitted and  
the incomplete gamma functions just replaced by half the corresponding
complete one. Furthermore, one can use for these complete gamma 
functions the Stirling asymptotic expansions $\Gamma(1+l)\sim
\exp[l\ln l-l+(1/2)\ln(2\pi l)]$ and the similar one for
$\Gamma(l+1+Y_2-Y_1)$. Neglecting terms of order $1/Y_1$ or smaller, for
a given value of $l-Y_1$, in $\ln(D/\gamma)$, one finds
\begin{equation} \label{4.18}
\frac{D(l+1,Y_1,Y_2)}{\gamma(l+1,Y_1)}=
\exp[(l-Y_1+\frac{1}{2})\ln(Y_2/Y_1)]
\end{equation}
Finally, one can shift the summation index in (\ref{4.15}), replacing 
$l-Y_1$ by $l-\bar{Y}_1$, where $\bar{Y}_1$ is the non-integer part of 
$Y_1$ such that $0\leq \bar{Y}_1<1$, and extend the summation to 
$-\infty$ in the present large-$Y_1$ limit. Thus, $\beta q^2=2$ being 
taken into account,
\begin{equation} \label{4.19}
\beta(\langle Q^2\rangle -\langle Q\rangle^2)=2\sum_{l=-\infty}^{\infty}
\frac{1}{\left(\alpha^{l-\bar{Y}_1+\frac{1}{2}}
+\alpha^{-(l-\bar{Y}_1+\frac{1}{2})}\right)^2}
\end{equation}
where $\alpha =R_2/R_1$. The form of (\ref{4.19}) justifies a posteriori 
our above statement that the relevant values of $l-Y_1$ in (\ref{4.15})
are only of the order of a few units. 

Here too, for a finite value of $\alpha$, the sum (\ref{4.19}) cannot be
replaced by an integral and does not reproduce the value (\ref{4.6})
expected on the basis of macroscopic electrostatics. The reason is the
same as in the case of the two-component plasma: the fluctuation
involves only a small number of particles. If $\bar{Y}_1=0$, i.e if the
background charge $-qn\pi R^2$ on the inner disk is an integer number of
elementary charges $-q$, the fluctuation (\ref{4.19}) for the
one-component plasma is the same as the fluctuation (\ref{4.9}) for the
two-component plasma; we have no explanation to offer.

The case of one disk alone, in the grand canonical ensemble (with a
fixed background), is obtained
by taking the limit $R_2\rightarrow\infty$. Now the average charge on
the disk has a simple form. For obtaining it, it is convenient to 
rewrite the summand in (\ref{4.13}) as $[1+(D/\gamma)]^{-1}$, where
now, from (\ref{4.18}), $D/\gamma =0$ if $l<Y_1-(1/2)$, 
$D/\gamma =1$ if $l=Y_1-(1/2)$, and $D/\gamma =\infty $ if
$l>Y_1-(1/2)$. Therefore, since the summation index $l$ is an integer,
\begin{eqnarray} 
&&\langle Q\rangle =-q\bar{Y}_1\quad {\mathrm if}\;\bar{Y}_1<\frac{1}{2} 
\nonumber \\
&&\langle Q\rangle =0\quad{\mathrm if}\;\bar{Y}_1=\frac{1}{2} \nonumber \\
&&\langle Q\rangle =q(1-\bar{Y}_1)\quad{\mathrm if}\;\bar{Y}_1>\frac{1}{2}
\label{4.20}
\end{eqnarray}
For the behavior of the fluctuation (\ref{4.19}), in the limit $\alpha
\rightarrow\infty$, two cases have to be distinguished.
\begin{eqnarray}
&&\langle Q^2\rangle-\langle Q\rangle^2=0\quad{\mathrm if}\;
\bar{Y}_1\neq\frac{1}{2} \nonumber \\
&&\langle Q^2\rangle-\langle Q\rangle^2=\frac{q^2}{4}\quad{\mathrm if}\; 
\bar{Y}_1=\frac{1}{2} \label{4.21}
\end{eqnarray}
These results have a simple interpretation: $-qY_1=-q\pi nR_1^2$ is the
negative background charge. Taking into account a (necessarily integer) 
number of positive particles gives a total charge (background plus
particles) $Q$. If $\bar{Y}_1\neq 1/2$, only one value of $Q$
contributes to the grand canonical ensemble: the one which corresponds
to the smallest possible value of $|Q|$. If however $\bar{Y}_1= 1/2$,
this smallest possible value is $|Q|=(q/2)$ which corresponds to 
two possibilities $Q=\pm (q/2)$ with equal probabilities. 

The two-dimensional ``plane capacitor'' limit ($R_1\rightarrow\infty$
for a fixed value of $W=R_2-R_1$) is the same as in the case of a
two-component plasma. As this limit is approached, the sum in
(\ref{4.19}) can be replaced by an integral, and $\langle Q\rangle = 0$
because of the geometrical symmetry between the two flat plates. Again,
the result is (\ref{4.10}), in agreement with macroscopic
electrostatics.
 
\renewcommand{\theequation}{5.\arabic{equation}}
\setcounter{equation}{0}

\section{\bf SUMMARY AND CONCLUSION}

\hspace*{\parindent} For studying the charge fluctuations on a
macroscopic but finite classical Coulomb system in the grand canonical
ensemble, i.e. when the system is allowed to exchange particles with a
reservoir, in a first step we have considered a capacitor: one electrode
is the finite system under consideration, the other electrode surrounds
the first one at some distance of it and extends to infinity. Both
electrodes are assumed to be made of the same Coulomb fluid. The
capacitor is short-circuited, which means that the two electrodes can
freely exchange particles. When the external electrode recedes to
infinity, the internal one becomes one Coulomb system in a grand
canonical ensemble. For actual calculations, the simple geometry of a
spherical capacitor has been chosen.

For a three-dimensional system, there is no surprise. A short-circuited
capacitor of capacitance $C$ can be considered as an electric
oscillator, and it is rather natural to state that its average energy
$\langle Q^2\rangle/2C$ is $(1/2)T$: this gives (\ref{2.1}). This
general formula is supported by the derivation of Section 2, on the
basis of linear response theory and macroscopic electrostatics. It has
been checked in Section 3, in the Debye-H\"uckel theory. 
 
Two-dimensional Coulomb systems (with a logarithmic interaction) are
more tricky. Since some of them are exactly solvable, it was tempting to
test on them the general formula (\ref{2.1}) for the charge
fluctuations. It has been a surprise for the author that this general
formula is \emph{not} valid for a circular capacitor at some finite
temperature. On second thought, one realizes that the charge
fluctuations involve only a small number of particles, and therefore 
one should not expect the validity of a macroscopic formula. 

In the limiting case of one disk alone in a grand canonical ensemble 
(i.e. the disk is  allowed to exchange particles with a reservoir at 
infinity), in general there is no charge fluctuation and the charge 
$Q$ is such that $|Q|$ has the smallest possible value (0 when possible, 
a fraction of the elementary charge $q$ in the case of a one-component 
plasma with a background charge which is not an integer number of 
elementary charges $-q$). An exception is when the background charge of
the one-component plasma is of the form$-[(N+(1/2)]q$ with $N$
integer. Then both values $Q=\pm(1/2)q$ are equally probable. We cannot
explain why there is no charge fluctuation in the cases when the
smallest possible value of the total charge is 0. Indeed, in these
cases, bringing another elementary charge $\pm q$ from infinity would
cost only a finite energy and one would expect $Q=\pm q$ to contribute
to the grand canonical ensemble. That these values $Q=\pm q$ do not
contribute might be a special feature of the solvable models at 
$\Gamma =2$. We just do not know.   

Another tricky feature of two-dimensional Coulomb systems is that the
two-dimensional Coulomb potential $-\ln(r/L)$ does not vanish at
infinity, if the length $L$ is finite. As discussed in Appendix B, for
obtaining sensible results in the Debye-H\"uckel theory, it is necessary
to take the limit $L\rightarrow\infty$.

\renewcommand{\theequation}{A.\arabic{equation}}
\setcounter{equation}{0}

\section*{\bf APPENDIX A: DEBYE-H\"UCKEL THEORY IN A SPHERICAL OR
CIRCULAR CAPACITOR}

\hspace*{\parindent} In the spherical capacitor geometry, the solution 
of (\ref{3.2}) can be expanded in Legendre 
polynomials $P_l(\cos\theta)$ (where $\theta$ is the angle between
${\mathbf r}$ and ${\mathbf r}'$) in the form
\begin{equation} \label{A.1}
K({\mathbf r},{\mathbf r}')=\sum_{l=0}^{\infty}k_l(r,r')P_l(\cos\theta)
\end{equation}
When this expansion is used in (\ref{3.1}) and (\ref{3.3}), only the
term $l=0$ survives the angular integrations. Thus, we only need the
function $k_0(r,r')$. The solution of (\ref{3.2}) for an infinite system
is $K({\mathbf r},{\mathbf r}')=-\beta\exp(|{\mathbf r}-{\mathbf r}'|)/
|{\mathbf r}-{\mathbf r}'|$ and its $l=0$ part is $-\beta
\sinh(\kappa r_<)\exp(-\kappa r_>)/(\kappa rr')$, where $r_<$ ($r_>$)
is the smaller (the larger) of $r$ and $r'$. In the 
present geometry, there are additional ``reflected'' and ``transmitted'' 
terms. When the source point ${\mathbf r}'$ is in the inner sphere
$(r'<R_1)$, the solution is of the form
\begin{eqnarray}
&&k_0(r,r')=-\frac{\beta}{\kappa rr'}[\sinh(\kappa r_<)\exp(-\kappa r_>)+
a\sinh(\kappa r)\sinh(\kappa r')]\quad (r,r'<R_1) \nonumber \\
&&k_0(r,r')=-\beta\frac{\sinh(\kappa r')}{r'}[\frac{b}{\kappa r}+c]
\quad (r'<R_1,\,R_1<r<R_2) \nonumber \\
&&k_0(r,r')=-\frac{\beta d}{\kappa rr'}\sinh(\kappa r')\exp(-\kappa r)
\quad (r'<R_1,\,r>R_2) \label{A.2}
\end{eqnarray}
This solution has the appropriate singularity at $r=r'$,
is otherwise regular at $r=0$ and $r'=0$, and goes to 0 when
$r\rightarrow\infty$. The four coefficients $a,b,c,d$ are to be
determined by the requirements that $k_0(r,r')$ and $\partial
k_0(r,r')/\partial r$ be continuous at $r=R_1$ and $r=R_2$. In
particular, one finds 
\begin{equation} \label{A.3}
a=\frac{\exp(-\kappa R_1)(\kappa R_2-\kappa R_1)}
{(1+\kappa R_2-\kappa R_1)\cosh (\kappa R_1)+\sinh (\kappa R_1)}
\end{equation}
Using the first equation (\ref{A.2}) and (\ref{A.3}) in (\ref{3.1}) and
(\ref{3.3}) gives (\ref{3.4}).

In two dimensions, in the circular capacitor geometry, the calculation
is very similar to the above one. The expansion (\ref{A.1}) is replaced
by
\begin{equation} \label{A.4}
K({\mathbf r},{\mathbf r}')=\sum_{l=0}^{\infty}k_l(r,r')\cos(l\theta)
\end{equation}
where the $l=0$ part is, in terms of modified Bessel functions $I_0$ and
$K_0$, when $r'<R_1$,
\begin{eqnarray}
&&k_0(r,r')=-\beta[I_0(\kappa r_<)K_0(\kappa r_>)
+aI_0(\kappa r)I_0(\kappa r')]\quad (r,r'<R_1) \nonumber \\
&&k_0(r,r')=-\beta I_0(\kappa r')[b\ln(\kappa r)+c]
\quad (r'<R_1,\,R_1<r<R_2) \nonumber \\
&&k_0(r,r')=-\beta dI_0(\kappa r')K_0(\kappa r)
\quad (r'<R_1,\,r>R_2) \label{A.5}
\end{eqnarray}
By the same method as above, the coefficient $a$ is found as
\begin{equation} \label{A.6}
a=\frac{-K_0(\kappa R_1)+\frac{R_1}{R_2}\frac{K_0(\kappa R_2)}
{K_1(\kappa R_2)}K_1(\kappa R_1)-\kappa R_1K_1(\kappa R_1)
\ln\frac{R_1}{R_2}}{I_0(\kappa R_1)+\frac{R_1}{R_2}
\frac{K_0(\kappa R_2)}{K_1(\kappa R_2)}I_1(\kappa R_1)
-\kappa R_1I_1(\kappa R_1)\ln\frac{R_1}{R_2}}
\end{equation}
Using the first equation (\ref{A.5}) and (\ref{A.6}) in (\ref{3.1}) 
and (\ref{3.3}), and the Wronskian of the Bessel functions, gives 
(\ref{4.2}).

\renewcommand{\theequation}{B.\arabic{equation}}
\setcounter{equation}{0}

\section*{\bf APPENDIX B: DEBYE-H\"UCKEL THEORY IN A FINITE SYSTEM}

\hspace*{\parindent} In Section 3 and Appendix A, the Debye-H\"uckel
differential equation (\ref{3.2}) was written and solved in the
spherical capacitor geometry; with minor modifications, the same
approach holds in the two-dimensional case, for a circular capacitor, 
as discussed in Section 4.2 and Appendix A. In these geometries, 
the Coulomb fluid extends to infinity in the region $r>R_2$ 
and it is obvious that the boundary condition should be 
$K({\mathbf r},{\mathbf r}')\rightarrow 0$ as $r\rightarrow\infty)$
(In this approach perfect screening is globally satisfied: The charge of
one particle plus the charge it induces in the two conductors sum to
zero. However, there is no perfect
screening if one takes into account only the charge of a particle
sitting on one of the conductors and the charge it induces on that
conductor only, and this gives rise to a charge fluctuation
on each conductor). The case of one sphere or disk in the grand 
canonical ensemble was obtained by taking the limit 
$R_2\rightarrow\infty$ at the end of the calculation. 

The question arises of how to formulate the Debye-H\"uckel 
theory in the grand canonical ensemble, directly starting with only a
sphere or a disk of radius $R$. What is the boundary condition to be
imposed at $r=R$ ? Choquard et al.\cite{Ch} have already investigated
this problem. Nevertheless, we revisit it, hoping to clarify some
delicate points.

We start with the three-dimensional case.
An unambiguous way of formulating the Debye-H\"uckel theory is to start
with a full diagrammatic expansion\cite{H} in the grand canonical
ensemble, and to make a topological reduction replacing the fugacity 
(fugacities) by the density (densities). The Debye-H\"uckel correlation 
function is obtained by resumming a class of diagrams (the chain
diagrams), or, equivalently, by taking for the function $K$ in
eq.(\ref{3.1}) the solution of the integral equation
\begin{equation} \label{B.1}
K({\mathbf r},{\mathbf r}')=-\frac{\beta}{|{\mathbf r}-{\mathbf r}'|}-
\frac{\kappa^2}{4\pi}\int{\mathrm d}{\mathbf r}''\frac{1}
{|{\mathbf r}-{\mathbf r}''|}K({\mathbf r}'',{\mathbf r}')
\end{equation}
This integral equation can also be seen as the Ornstein-Zernicke
equation in which the direct correlation function between particles of
species $a$ and $b$ is approximated by $-\beta$ times their bare Coulomb
interaction $q_aq_b/|{\mathbf r}-{\mathbf r}'|$. In a finite system, the 
densities $n_a$ in $\kappa^2=4\pi\beta\sum_a n_aq_a^2$ are 
position-dependent near the boundaries. However, for the large-size 
systems considered here, this effect can be neglected and $\kappa^2$ 
will be taken as a constant in the whole system. 

By taking the Laplacian of both sides of the integral equation
(\ref{B.1}), one obtains the partial differential equation (\ref{3.2}). 
However, the integral equation provides the boundary condition to be
used in (\ref{3.2}). In the presently studied case of a sphere
of radius $R$,  
we can use the expansion (\ref{A.1}) in Legendre polynomials. For 
brevity we only consider the $l=0$ part. The integral equation
(\ref{B.1}) gives for the boundary condition on the surface $r=R$ 
(with $r'<R$)
\begin{equation} \label{B.2}
k_0(R,r')=-\frac{\beta}{R}[1+\frac{\kappa^2}{\beta}\int_0^R
{\mathrm d}r''\,r''^2k_0(r'',r')]
\end{equation}
Using in (\ref{B.2}) the general form (\ref{A.2}) of the solution of 
the partial differential equation (\ref{3.2}) (with now $r,r'\leq R$) and
performing the integral determines the coefficient $a$ as
\begin{equation} \label{B.3}
a=\frac{\exp(-\kappa R)}{\cosh(\kappa R)}
\end {equation}   
(The square bracket in the r.h.s. of (\ref{B.2}) does not vanish: there 
is no perfect screening on the sole sphere).
This directly obtained result (\ref{B.3}) is identical to the limit of
the spherical capacitor $a$ coefficient (\ref{A.3}) (with $R_1=R$) as 
$R_2\rightarrow\infty$. One recovers the mean square charge (\ref{2.4}).
The existence of a fluctuation confirms that the formulation of the 
Debye-H\"uckel theory by the integral equation (\ref{B.1}) is a grand
canonical one. 

The two-dimensional case is more tricky. The bare Coulomb interaction
between two unit charges now is $-\ln(|{\mathbf r}-{\mathbf r}'|/L)$,
with $L$ some fixed length. The integral equation now is
\begin{equation} \label{B.4}
K({\mathbf r},{\mathbf r}')=\beta\ln\frac{|{\mathbf r}-{\mathbf r}'|}{L}
+\frac{\kappa^2}{2\pi}\int{\mathrm d}{\mathbf r}''\ln\frac
{|{\mathbf r}-{\mathbf r}''|}{L}K({\mathbf r}'',{\mathbf r}')
\end{equation}
with $\kappa^2=2\pi\beta\sum_an_aq_a^2$. The $l=0$ part of the expansion
(\ref{A.4}) obeys the boundary condition (with $r'<R$)
\begin{equation} \label{B.5}
k_0(R,r')=\beta\ln\frac{R}{L}[1+\frac{\kappa^2}{\beta}\int_0^R  
{\mathrm d}r''\,r''k_0(r'',r')]
\end{equation}
Using in (\ref{B.5}) the general form (\ref{A.5}) of the solution of
the two-dimensional analog of  
the partial differential equation (\ref{3.2}) (with now $r,r'\leq R$)
and performing the integral determines the coefficient $a$ as
\begin{equation} \label{B.6}
a=\frac{\kappa RK_1(\kappa R)\ln\frac{R}{L}+K_0(\kappa R)}
{\kappa RI_1(\kappa R)\ln\frac{R}{L}-I_0(\kappa R)}
\end{equation}

This result (\ref{B.6}) is unacceptable as it stands, since it depends
on the arbitrary length $L$, which only determines the zero of the 
potential and should not enter the correlation functions. Actually,
using a Coulomb interaction $-\ln(r/L)$ which does not vanish at
infinity causes difficulties at many places. For instance, the Coulomb
energy of a macroscopic disk of radius $R$, carrying the macroscopic 
charge $Q$ near its circumference, would be $(1/2)Q^2\ln(L/R)$, negative 
if $R>L$. Thus, in the grand canonical ensemble, configurations of
infinite $|Q|$ would be favored, causing the grand canonical partition
function to diverge. This seems to indicate that the limit
$L\rightarrow\infty$ (such that the zero of the potential recedes to
infinity) should be taken in (\ref{B.6}), which then becomes
\begin{equation} \label{B.7}
a=\frac{K_1(\kappa R)}{I_1(\kappa R)}
\end{equation}
This result (\ref{B.7}) is identical to the limit of the circular
capacitor $a$ coefficient (\ref{A.6}) (with $R_1=R$) as
$R_2\rightarrow\infty$. It might even be noted that, if in (\ref{A.6})
we take $R_1=R$ and $R_2=L\gg R$ and neglect the term of order $R_1/R_2=
R/L$ in both the numerator and the denominator, (\ref{B.6}) is
recovered. This is a further indication that the limit 
$L\rightarrow\infty$ should be taken in the case of a system made of one
disk only. Now, there is perfect screening (the square bracket in the 
r.h.s. of (\ref{B.5}) vanishes) and there is no charge fluctuation on 
the disk. 
 
Still another way of dealing with a finite system, in three or two
dimensions,  would be to first assume that the whole space external to
the system is filled with a medium of Debye wave number $\kappa'$ and
solve the partial differential equation (\ref{3.2}) (or its
two-dimensional analog) in the whole space, taking 
$\kappa^2({\mathbf r})=\kappa^2$ in the system and
$\kappa^2({\mathbf r})=\kappa'^2$ outside, with the proper continuity
conditions on the system boundary and $K({\mathbf r},{\mathbf r}')
\rightarrow 0$ as $r\rightarrow\infty$. One recovers the same results as
above.

\renewcommand{\theequation}{C.\arabic{equation}}
\setcounter{equation}{0}

\section*{{\bf APPENDIX C: TWO-DIMENSIONAL \\ 
TWO-COMPONENT PLASMA AT} ${\mathbf \Gamma=2}$}

\hspace*{\parindent} For this exactly solvable model\cite{CJ2}, the 
charge correlation function can be expressed in terms of Green 
functions $G_{++}({\mathbf r},{\mathbf r}')$ and 
$G_{-+}({\mathbf r},{\mathbf r}')$ as
\begin{equation} \label{C.1}
\langle\rho({\mathbf r})\rho({\mathbf r}')\rangle=
-2m^2q^2[|G_{++}({\mathbf r},{\mathbf r}')|^2+
|G_{-+}({\mathbf r},{\mathbf r}')|^2]+
n({\mathbf r})q^2\delta({\mathbf r}-{\mathbf r}')
\end{equation}
where $m$ is a rescaled fugacity such that $(2m)^{-1}$ is the bulk
correlation length, and $n({\mathbf r})$ is the total number density 
(actually, $n$ is a divergent quantity, but it will be seen that
this divergence causes no difficulty here). In (\ref{C.1}), the charge 
symmetry of the model has been taken into account. We consider the 
circular capacitor geometry and the source point ${\mathbf r}'$ is in
the inner disk ($r'<R_1$). For (\ref{C.1}) to be the charge correlation
function, ${\mathbf r}$ must be in a filled region ($r<R_1$ or
$r>R_2$). 

When ${\mathbf r}$ is in a filled region, the Green functions obey the
equations\cite{CJ2} 
\begin{equation} \label{C.2}  
(m^2-\Delta)G_{++}({\mathbf r},{\mathbf r}')=
m\delta({\mathbf r}-{\mathbf r}')
\end{equation}
and
\begin{equation} \label{C.3}
G_{-+}({\mathbf r},{\mathbf r}')=-\frac{\exp({\mathrm i}\varphi)}
{m}\left(\frac{\partial}{\partial r}+\frac{{\mathrm i}}{r}\frac
{\partial}{\partial\varphi}\right)G_{++}({\mathbf r},{\mathbf r}')
\end{equation}
where $(r,\varphi)$ are the polar coordinates of ${\mathbf r}$.
The solution of (\ref{C.2}) is an expansion of the form
\begin{eqnarray} 
&&G_{++}({\mathbf r},{\mathbf r}')=
\frac{m}{2\pi}\sum_{l=-\infty}^{\infty}\left[I_l(mr_<)K_l(mr_>)
+a_lI_l(mr')I_l(mr)\right]\exp[{\mathrm i}l(\varphi-\varphi')]
\quad (r,r'<R_1) \nonumber \\
&&G_{++}({\mathbf r},{\mathbf r}')=
\frac{m}{2\pi}\sum_{l=-\infty}^{\infty}d_lI_l(mr')K_l(mr)
\exp[{\mathrm i}l(\varphi-\varphi')]\quad (r'<R_1,\,r>R_2)
\label{C.4}
\end{eqnarray}
These expansions have the proper singularity at 
${\mathbf r}={\mathbf r}'$, are otherwise regular at $r=0$ and $r'=0$,
and go to zero when $r\rightarrow\infty$. As to $G_{-+}$, (\ref{C.3}) 
gives
\begin{eqnarray}
&&G_{-+}({\mathbf r},{\mathbf r}')= \nonumber \\
&&\frac{m}{2\pi}\sum_{l=-\infty}^{\infty}\left[I_l(mr')K_{l+1}(mr)
-a_lI_l(mr')I_{l+1}(mr)\right]\exp[{\mathrm i}(l+1)\varphi-
{\mathrm i}l\varphi']\quad (r'<r<R_1) \nonumber \\
&&G_{-+}({\mathbf r},{\mathbf r}')= \nonumber \\
&&\frac{m}{2\pi}\sum_{l=-\infty}^{\infty}d_lI_l(mr')K_{l+1}(mr)
\exp[{\mathrm i}(l+1)\varphi-{\mathrm i}l\varphi']
\quad (r'<R_1,\,r>R_2) \label{C.5}
\end{eqnarray}

When ${\mathbf r}$ is in the empty region $R_1<r<R_2$ (where $m=0$), 
as functions of ${\mathbf r}$, $G_{++}$ depends only on 
$z=r\exp({\mathrm i}\varphi)$ and $G_{-+}$ depends only on 
$\bar{z}=r\exp(-{\mathrm i}\varphi)$. Thus, the expansions are of the
forms
\begin{equation} \label{C.6}
G_{++}({\mathbf r},{\mathbf r}')=
\frac{m}{2\pi}\sum_{l=-\infty}^{\infty}b_lI_l(mr')(mr)^l
\exp[{\mathrm i}l(\varphi-\varphi')]\quad (r'<R_1,\,R_1<r<R_2)
\end{equation}
and
\begin{equation} \label{C.7}
G_{-+}({\mathbf r},{\mathbf r}')=
\frac{m}{2\pi}\sum_{l=-\infty}^{\infty}c_lI_l(mr')(mr)^{-(l+1)}
\exp[{\mathrm i}(l+1)\varphi-{\mathrm i}l\varphi']
\quad (r'<R_1,\,R_1<r<R_2)
\end{equation}

The coefficients $a_l,b_l,c_l,d_l$ are to be determined by the 
requirements that $G_{++}$ and $G_{-+}$ be continuous at $r=R_1$ 
and $r=R_2$. In particular, after having used some functional 
relations between Bessel functions, one finds
\begin{equation} \label{C.8}
d_l=\left[lI_l(X)K_l(\alpha X)(\alpha^l-\alpha^{-l})
-XI_l(X)K'_l(\alpha X)\alpha^{l+1}
+XI'_l(X)K_l(\alpha X)\alpha^{-l}\right]^{-1}
\end{equation}
where $X=mR_1$ and $\alpha=R_2/R_1$.

For computing the mean square charge on the inner disk, using
the perfect screening relation 
$\int{\mathrm d}{\mathbf r}\langle\rho ({\mathbf r})
\rho ({\mathbf r'})\rangle =0$, where the integral is on the whole
space, it is convenient to rewrite (\ref{3.3}) as
\begin{equation} \label{C.9} 
\langle Q^2\rangle =-\int_{r>R_2}{\mathrm d}{\mathbf r}\int_{r'<R_1}
{\mathrm d}{\mathbf r'}\langle\rho ({\mathbf r})\rho 
({\mathbf r'})\rangle
\end{equation}
and to use (\ref{C.1}), omitting the self part $nq^2\delta({\mathbf r}-
{\mathbf r}')$ which does not contribute to (\ref{C.9}). The
angular integrations are easily performed, by using the mutual 
orthogonality of the functions $\exp({\mathrm i}l\varphi)$, and 
(\ref{C.9}) becomes
\begin{equation} \label{C.10}
\langle Q^2\rangle =2m^4q^2\sum_{l=-\infty}^{\infty}d_l^2
\int_0^{R_1}{\mathrm d}r'\,r'I_l^2(mr')
\int_{R_2}^{\infty}{\mathrm d}r\,r[K_l^2(mr)+K_{l+1}^2(mr)]
\end{equation}
where $d_l$ is given by (\ref{C.8}). After having performed the 
integrations in (\ref{C.10}) and used some functional relations 
between Bessel functions, one obtains (\ref{4.7}).

\renewcommand{\theequation}{D.\arabic{equation}}
\setcounter{equation}{0}

\section*{{\bf APPENDIX D: TWO-DIMENSIONAL \\
ONE-COMPONENT PLASMA AT} ${\mathbf \Gamma=2}$}

\hspace*{\parindent} For this exactly solvable model\cite{BJ1,CJB}, 
the charge density (including the background contribution) and the
charge correlation function can be expressed in terms of a projector
$P({\mathbf r},{\mathbf r'})$  
\begin{equation} \label{D.1}
\langle\rho ({\mathbf r})\rangle=q[P({\mathbf r},{\mathbf r})-n]
\end{equation}
and
\begin{equation} \label{D.2}
\langle\rho ({\mathbf r})\rho({\mathbf r'})\rangle-
\langle\rho ({\mathbf r})\rangle\langle\rho({\mathbf r'})\rangle=
q^2[-|P({\mathbf r},{\mathbf r'})|^2+P({\mathbf r},{\mathbf r})
\delta({\mathbf r}-{\mathbf r'})]
\end{equation}
We consider the circular capacitor geometry. For ({D.1}) and ({D.2}) to
be the charge density and correlation, respectively, ${\mathbf r}$ and
${\mathbf r'}$ must be in a filled region.

The electric potential $qV(r)$ created by the background
will be needed. It obeys $\Delta V(r)=2\pi n$ in the filled regions 
$r<R_1$ and $r>R_2$. It obeys $\Delta V(r) =0$ in the empty region 
$R_1<r<R_2$. At the boundaries $r=R_1$ and $r=R_2$, $V(r)$ and 
${\mathrm d}V/{\mathrm d}r$ must be continuous. Up to an overall 
irrelevant additive constant,
\begin{eqnarray} 
&&V(r)=\frac{1}{2}\pi nr^2\quad (r<R_1) \nonumber \\
&&V(r)=\frac{1}{2}\pi nR_1^2+\pi nR_1^2\ln\frac{r}{R_1}\quad 
(R_1<r<R_2) \nonumber\\
&&V(r)=\frac{1}{2}\pi n(R_1^2-R_2^2)+\pi nR_1^2\ln\frac{R_2}{R_1}+
\nonumber \\ 
&&\frac{1}{2}\pi nr^2+\frac{1}{2}\pi n(R_1^2-R_2^2)\ln\frac{r}{R_2}
\quad (r>R_2) \label{D.3}
\end{eqnarray}

$P({\mathbf r},{\mathbf r'})$ is the projector on the functional space
spanned by the functions $\psi_l({\mathbf r})=
\exp[-V(r)][r\exp({\mathrm i}\varphi)]^l\;(l=0,1,2,3,\ldots)$ (this
definition of $\psi_l$ holds in the filled regions $r<R_1$ and $r>R_2$,
while $\psi_l({\mathbf r})=0$ in the empty region $R_1<r<R_2$):
\begin{equation} \label{D.4}
P({\mathbf r},{\mathbf r'})=\sum_{l=0}^{\infty}\frac{1}{C_l}
\psi_l({\mathbf r})\bar{\psi}_l({\mathbf r'})=
\sum_{l=0}^{\infty}\frac{1}{C_l}\exp[-V(r)-V(r')]r^lr'^l\exp[{\mathrm i}
l(\varphi-\varphi')]
\end{equation}
where $C_l$ is the normalization constant
\begin{eqnarray} 
C_l&=&(\int_{r<R_1}+\int_{r>R_2}){\mathrm d}{\mathbf r}\,
\exp[-2V(r)]r^{2l} \nonumber \\
&=&\frac{\pi}{(\pi n)^{l+1}}[\gamma(l+1,Y_1)+D(l+1,Y_1,Y_2)] \label{D.5}
\end{eqnarray}
where the functions $\gamma$ and $D$ are defined in (\ref{4.11}) and
(\ref{4.14}). 

The average charge on the inner disk is  
\begin{equation} \label{D.6}
\langle Q\rangle =\int_{r<R_1}{\mathrm d}{\mathbf r}\langle\rho 
({\mathbf r})\rangle
\end{equation}
Using (\ref{D.1}) and (\ref{D.4}) in (\ref{D.6}) and performing the
integration gives (\ref{4.13}).

The charge fluctuation on the inner disk is
\begin{equation} \label{D.7}
\langle Q^2\rangle -\langle Q\rangle ^2=
\int_{r<R_1}{\mathrm d}{\mathbf r}\int_{r'<R_1}{\mathrm d}{\mathbf r}'
[\langle\rho ({\mathbf r})\rho({\mathbf r'})\rangle-
\langle\rho ({\mathbf r})\rangle\langle\rho({\mathbf r'})\rangle]
\end{equation}
Using (\ref{D.2}) and (\ref{D.4}) in (\ref{D.7}) and performing the
integrations, using the mutual orthogonality of the functions 
$\exp({\mathrm i}l\varphi)$, gives (\ref{4.15}).

\section*{\bf ACKNOWLEDGEMENTS}

\hspace*{\parindent} I have benefited of stimulating discussions 
with J. L. Lebowitz, L. \v{S}amaj, F. Cornu, E. Trizac, and many 
others.

\end{document}